\documentclass[a4paper]{jpconf}
\usepackage{graphicx}
\begin{document}
\title{Theoretical study on isotopic shift in angle-resolved photoemission
spectra of Bi$_2$Sr$_2$CaCu$_2$O$_8$}

\author{Kai Ji and Keiichiro Nasu}

\address{
CREST JST, Solid State Theory Division,
Institute of Materials Structure Science, KEK,
Graduate University for Advanced Studies,
Oho 1-1, Tsukuba, Ibaraki 305-0801, Japan
}

\ead{jikai@post.kek.jp}

\begin{abstract}
We develop a path-integral theory to study the angle-resolved photoemission
spectra (ARPES) of high-$T_c$ superconductors based on a two-dimensional
model for the CuO$_2$ conduction plane, including both electron-electron
($e$-$e$) and electron-phonon ($e$-ph) interactions.
Comparing our result with the experimental one of Bi$_2$Sr$_2$CaCu$_2$O$_8$,
we find that the experimentally observed isotopic band shift in ARPES
is due to the off-diagonal quadratic $e$-ph coupling, whereas the presence
of $e$-$e$ repulsion partially suppresses this effect.
\end{abstract}

\section{Introduction}

The study of high-$T_c$ superconductivity is one of the most attractive
realms in the last two decades.
Since the angle-resolved photoemission spectroscopy (ARPES) directly probes
the electronic occupied states, it has become an important technique to
investigate the electronic properties of cuprates\cite{da03}.
Recently, the oxygen isotope effect has been detected with ARPES in
Bi$_2$Sr$_2$CaCu$_2$O$_8$ (Bi2212) by two groups\cite{gw04,do07},
and a common feature is noticed that the spectra are shifted with the
$^{16}$O/$^{18}$O substitution, providing direct evidence for electron-phonon
($e$-ph) coupling in this material.
However, since the first report by Gweon {\it et al.}\cite{gw04}, this
isotopic band shift has become an controversial issue\cite{ma05},
as the observed shift is up to 40 meV, much larger than the isotopic energy
change of oxygen phonon, $\sim$ 5 meV\cite{ma95}.
Very recently, Douglas {\it et al.}\cite{do07} repeat the experiment, but
find the shift is only 2$\pm$3 meV.
Thus it turns out to be an interesting problem whether the large shift 
observed by Gweon {\it et al.} is possible or not in the cuprates.
In this paper we shall look into this isotope induced band shift from
a theoretical point of view.

In the CuO$_2$ plane of cuprates, as shown in Fig. 1, the electronic transfer
is affected by the vibration of oxygen atoms between the initial and final
Cu sites, resulting in an off-diagonal type $e$-ph coupling.
In order to have an insight into the isotope effect of Bi2212, we start
from a half-filled Hamiltonian including the electron-electron ($e$-$e$)
repulsion and the above mentioned off-diagonal $e$-ph coupling
($\hbar = 1$ and $k_B = 1$ throughout this paper):
\begin{eqnarray}
H &=& - \sum_{\langle l,l' \rangle, \sigma} t (l, l')
  (a^{\dag}_{l \sigma} a_{l' \sigma} + a^{\dag}_{l' \sigma} a_{l \sigma})
  +U \sum_l n_{l \uparrow} n_{l \downarrow}
  + {\omega_0 \over 2} \sum_{\langle l,l' \rangle} \left(-{1 \over \lambda}
  \frac{\partial^2}{\partial q^2_{ll'}} + q^2_{ll'} \right),
\end{eqnarray}
where $a^{\dag}_{l \sigma}$ ($a_{l \sigma}$) is the creation (annihilation)
operator of an electron with spin $\sigma$ at the Cu site $l$ on a square
lattice (see in Fig. 1).
The electrons hop between two nearest neighboring Cu sites, denoted by
$\langle l, l' \rangle$, with a transfer energy $t(l, l')$.
$U$ is the strength of Coulomb repulsion between two electrons on the
same Cu site with opposite spins.
The oxygen phonon is assumed to be of the Einstein type with a frequency
$\omega_0$ and a mass $m$.
$\lambda$ ($\equiv 1 + \Delta m / m$)
is the mass change factor of phonon due to the isotope substitution.
In the third term, $q_{ll'}$ is the dimensionless coordinate operator
of the oxygen phonon locating between the nearest-neighboring Cu sites
$l$ and $l'$, and the sum denoted by ${\langle l,l' \rangle}$ just means
a summation over all the phonon sites in the lattice.

\begin{figure}[h] 
\begin{center}
\includegraphics[width=8pc]{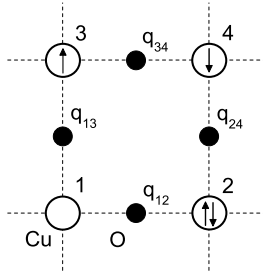}
\hspace{2pc}
\begin{minipage}[b]{20pc}
\caption{A Schematic plot of CuO$_2$ conduction plane in cuprates.
The copper atom (white circle) is on the $l$-th site of a simple square lattice.
The oxygen atom (black circle) is located between the nearest-neighboring Cu
sites, and $q_{l l'}$ denotes its displacement from the equilibrium position.}
\end{minipage} 
\end{center}
\end{figure}

In the conduction plane of CuO$_2$, the electronic hopping integral
$t (l, l')$ can be expanded to the second order terms with respect to
the phonon displacements $q_{ll'}$ as
\begin{equation}
t (l, l') = t_0 + s q^2_{ll'},
\end{equation}
where $t_0$ is the bare hopping energy and $s$ the off-diagonal quadratic
$e$-ph coupling constant.
Here we note the linear $e$-ph coupling does not occur owing to the lattice
symmetry of present model.
Whereas the inter-site $e$-$e$ interaction is included in the screened
values of $t_0$ and $s$.

\section{Path-integral Monte Carlo}

In this section, we develop a path-integral theory for a model with both
$e$-$e$ and $e$-ph interactions.
By making use of the Trotter's decoupling formula, the Boltzmann operator
is written as,
\begin{eqnarray}
e^{-\beta H} = \lim_{L \rightarrow \infty}
	\left( e^{-\Delta H} \cdots e^{-\Delta H} \right).
\end{eqnarray}
Applying the Hubbard-Stratonovitch transformation\cite{to97} and the Gaussian
integral formula\cite{ji04}, we decouple the two-body parts, so that the
$e$-$e$ and $e$-ph correlated terms are replaced by a two-fold summation
over the auxiliary spin and lattice configurations, which is the so-called
path-integral.
In this way, the Boltzmann operator is rewritten into the path-integral
form as,
\begin{eqnarray}
e^{-\beta H} & \rightarrow & \int {\mathcal D} x
  \left( T_+ \exp \left\{ -\int_0^\beta d \tau
  \left[ h(\tau,x_m,x_q) + \Omega(x_q) \right] \right\}
  \prod_l \left[ |x_q(l,\theta) \rangle \langle x_q(l,0)|
  \right] \right),
\end{eqnarray}
\begin{eqnarray}
h(\tau,x_m,x_q) & \equiv & - \sum_{\langle l,l' \rangle, \sigma}
  \left[ t_0 + s x^2_q(l,l',\tau) \right]
  \left[a^{\dag}_{l \sigma}(\tau) a_{l' \sigma}(\tau)
  + a^{\dag}_{l' \sigma}(\tau) a_{l \sigma}(\tau) \right]
  \nonumber\\
& &	- \sqrt{U \over \Delta} \sum_l x_m(l, \tau)
  [ n_{l \uparrow}(\tau) - n_{l \downarrow}(\tau) ],
  \\
  \Omega(x_q) & \equiv & \sum_{\langle l,l' \rangle}
  \left\{ {\lambda \over 2 \omega_0}
  \left[ {\partial x_q(l,l',\tau) \over \partial \tau} \right]^2
  + {1 \over 2} \omega_0 x^2_q(l,l',\tau) \right\}.
\end{eqnarray}
Here, $x_m$ and $x_q$ correspond to the auxiliary spin and lattice field,
respectively, and $\int {\mathcal D} x$ symbolically denotes the integrals
over the path $x$ synthesized by $x_m$ and $x_q$.
$\Delta$ is the time interval of the Trotter's formula,
$\beta \equiv 1/T$, and $T$ is the absolute temperature.
$T_+$ in Eq. (4) is the time ordering operator.

Then we define the time evolution operator [$\equiv R (\tau, x)$] as
\begin{eqnarray}
R(\tau,x) = T_+ \exp \left[ -\int_0^{\tau} d \tau'
	h ( \tau', x_m, x_q ) \right].
\end{eqnarray}
In terms of the Boltzmann operator (4) and time evolution operator (7),
we define the free energy [$\equiv \Phi (x)$] of the given path as
\begin{eqnarray}
e^{-\beta \Phi (x)} = e^{-\int_0^{\beta} d \tau \Omega(x_q)}
	{\rm Tr} \left[ R(\beta, x) \right].
\end{eqnarray}
While, the partition function ($\equiv Z$) and total free energy
($\equiv \Phi$) are given as
\begin{eqnarray}
Z = e^{-\beta \Phi} = \int {\mathcal D}x
  e^{-\beta \Phi (x)}.
\end{eqnarray}
According to Refs. [6] and [7], we also define the one-body Green's function
[$\equiv G_{\sigma} (l \tau, l' \tau', x)$] on a path $x$ as
\begin{eqnarray}
G_{\sigma} (l \tau, l' \tau', x) = - \mbox{sign} (\tau - \tau')
	\langle T_+ \vec{a}_{l \sigma} (\tau)
	\vec{a}^{\dag}_{l' \sigma} (\tau') \rangle_x,
\end{eqnarray}
where $\vec{a}_{l \sigma} (\tau)$ is the Heisenberg representation of
$a_{l \sigma}$.
It is really time-dependent and defined by
\begin{eqnarray}
\vec{a}_{l \sigma} (\tau) \equiv R^{-1}(\tau,x) a_{l \sigma} R(\tau,x) .
\end{eqnarray}
Meanwhile, the ordinary Green's function [$\equiv G_{\sigma} (l, \tau)$]
can be obtained by the path-integral as
\begin{equation}
G_{\sigma} (l - l', \tau - \tau') = {1 \over Z} \int
	{\mathcal D}x e^{- \beta \Phi (x)}
	G_{\sigma} (l \tau, l' \tau', x).
\end{equation}
This path-integral is evaluated by the quantum Monte Carlo (QMC) simulation
method.

If the QMC data of Green's function $G_{\sigma} (l, \tau)$ is obtained,
we can immediately calculate its Fourier component
[$\equiv G_{\sigma} ({\bf k}, \tau)$] as
\begin{eqnarray}
G_{\sigma} ({\bf k}, \tau) = {1 \over N} \sum_l
	G_{\sigma} (l, \tau) e^{-i {\bf k} \cdot {\bf R}_l} ,
\end{eqnarray}
where $\bf k$ is the momentum of the outgoing photo-electron.
From this Fourier component $G_{\sigma} ({\bf k}, \tau)$, we derive the
spectral function [$\equiv A_{\sigma} ({\bf k}, \omega)$] by solving the
integral equation
\begin{eqnarray}
G_{\sigma} ({\bf k}, \tau) = - \int^{\infty}_{- \infty} d \omega
	\frac {e^{- \tau \omega}} {1 + e^{- \beta \omega}}
	A_{\sigma} ({\bf k}, \omega) .
\end{eqnarray}
Finally, the normalized spectral intensity is obtained as,
\begin{eqnarray}
I ({\bf k}, \omega) =
  \frac{\sum_{\sigma} A_{\sigma} ({\bf k}, \omega) f (\omega)}
  {\int d \omega \sum_{\sigma} A_{\sigma} ({\bf k}, \omega) f (\omega)} ,
\end{eqnarray}
where the Fermi-Dirac function $f (\omega) = 1 / [\exp(\beta \omega) + 1]$
is imposed.

\section{Results and Discussions}

We now present the QMC results on a 4$\times$4 square lattice, where $t_0$
is set as the unit of energy, and $\omega_0$=1.0 is used.
For the QMC simulation, we impose a little large isotopic mass enhancement,
$\lambda_0$=1 and $\lambda$=2, to suppress the numerical error.
In this calculation, we determine the binding energy $\epsilon_{\bf k}$
by the moment analysis of the spectral intensity as
$\epsilon_{\bf k} = \int d \omega I({\bf k}, \omega) \omega$.
Correspondingly, the isotope induced band shift is calculated by
$\Delta \epsilon_{\bf k} \equiv
\epsilon_{\bf k} (\lambda) - \epsilon_{\bf k} (\lambda_0)$.

\begin{figure}[h] 
\begin{center}
\includegraphics[width=26pc]{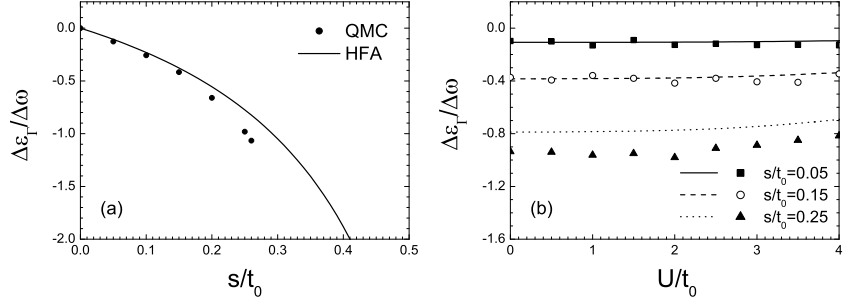}
\begin{minipage}{36pc}
\caption{
(a) The variation of $\Delta \epsilon_{\Gamma} / \Delta \omega$ with $s$
on a 4$\times$4 square lattice, when $U$=2.0, $\beta$=10, $\lambda_0$=1,
$\lambda$=2.
The filled circles are from QMC, and the solid curve from HFA as a guide
for eyes.
(b) The variations of $\Delta \epsilon_{\Gamma} / \Delta \omega$ with $U$
on a 4$\times$4 square lattice at $\beta$=10, $\lambda_0$=1 and $\lambda$=2.
Three different values of $s$ are used to show the $s$-dependence of
$\Delta \epsilon_{\Gamma} / \Delta \omega$.
The discrete symbols are the results of QMC, and continuous curves by
HFA as a reference.
}
\label{f2}
\end{minipage} 
\end{center}
\end{figure}

In Fig. 2(a), we plot the ratio $\Delta \epsilon_{\Gamma} / \Delta \omega$
versus $s$, at $U$=2.0 and $\beta$=10, where $\Delta \epsilon_{\Gamma}$ is
the band shift at the $\Gamma$ point of Brillouin zone
[${\bf k}_{\Gamma}$=(0,0)], and $\Delta \omega$ is the isotopic change
of phonon energy.
The filled circles are calculated by QMC, and the solid curve by the
mean-field theory with Hartree-Fork approximation (HFA) as a guide for
eyes.
Here both theories figure out an increase of
$\Delta \epsilon_{\Gamma} / \Delta \omega$ with $s$, which means if the
$e$-ph coupling is strong enough, a large band shift can be generated
in the cost of a small $\Delta \omega$.
In Fig. 2(b), the ratio $\Delta \epsilon_{\Gamma} / \Delta \omega$ versus
$U$ are shown for three different $s$'s, where the discrete symbols and
continuous curves are the QMC and HFA results, respectively.
It can be seen that the ratio $\Delta \epsilon_{\Gamma} / \Delta \omega$
increases with $s$.
Meanwhile, for a fixed $s$, the ratio declines slightly as $U$ increases.
This behavior indicates that the band shift is owing to the $e$-ph coupling,
whereas the presence of $U$ partially reduces this effect.
In terms of Figs. 2(a) and 2(b), one can see the band shift is actually
a measure of the $e$-ph coupling strength in the system.
If the result of Ref. [2] is correct, the $e$-ph coupling must be strong
in Bi2212.
On the contrary, Ref. [3] shows that the coupling cannot be very large.

\section{Conclusion}

In summary, by using the path-integral QMC method, we study the isotopic
shift in the ARPES of Bi2212 based on a model including both $e$-$e$
and off-diagonal quadratic $e$-ph interactions.
Our calculation demonstrates that the band shift is primarily triggered
by the $e$-ph coupling, while the presence of $e$-$e$ repulsion tends
to suppress this effect.

\end{document}